\documentclass[prl,aps,twocolumn,showpacs]{revtex4}

\usepackage{epsfig}

\begin{document}

\title{The Absence of Superconductivity in Single Phase CaFe$_2$As$_2$ under Hydrostatic Pressure}

\author{W. Yu$^{1,3}$}
\email{yuweiq@mcmaster.ca}
\author{A. A. Aczel$^{1}$}
\author{T. J. Williams $^{1}$}
\author{S. L. Bud'ko$^{2}$}
\author{N. Ni$^{2}$}
\author{P. C. Canfield$^{2}$}
\author{G. M. Luke$^{1,4}$}
\affiliation{$^1$Department of Physics and Astronomy, McMaster University, Hamilton, Ontario L8S 4M1, Canada\\
$^2$Department of Physics and Astronomy and Ames Laboratory, Iowa State University, Ames, Iowa 50011, USA\\
$^3$Department of Physics, Renmin University of China, Beijing 100872, China\\
$^4$Canadian Institute for Advanced Research, Toronto, Canada}

\date{\today}

\begin{abstract}

Recent high-pressure studies found that superconductivity can be achieved under very low pressure in the parent iron arsenide compound CaFe$_2$As$_2$, although
details of the sharpness and temperature of transitions vary between liquid medium and gas medium measurements.
To better understand this issue, we performed high-pressure susceptibility and transport studies on 
CaFe$_2$As$_2$, using helium as the pressure medium. The signatures of the transitions to the low-temperature orthorhombic and collapsed tetragonal
phase remained exceptionally sharp and no signature of bulk superconductivity was found under our hydrostatic 
conditions. Our results suggest that phase separation and superconductivity in CaFe$_2$As$_2$ are induced by non-hydrostatic conditions associated with the frozen liquid
media.

\end{abstract}

\maketitle

The recent discovery of superconductivity in doped iron arsenide compounds\cite{Hosono_Jacs1, Takahashi_Nature,Rotter_PRL_101_107006} and the later improvement of the superconducting 
transition temperature T$_c$ in both the pnictide oxides such as ROFeAs (R111) \cite{Chen_XH,Chen_GF,Ren_ZA, Wen_HH} and the ThCr$_2$Si$_2$-structure compounds such as 
(Ba,K)Fe$_2$As$_2$ (R122) \cite{Rotter_PRL_101_107006} have caused extensive experimental and theoretical studies in this new class of materials with
layered FeAs planes. Similar to the high-T$_c$ cuprates, the parent compounds exhibit structural transitions from a high-temperature tetragonal 
phase to a low-temperature orthorhombic phase, and the orthorhombic phase is usually antiferromagnetically (AF) ordered\cite{Dai_PC}. Upon doping, 
both the orthorhombic structure and the AF phase are suppressed and superconductivity is induced.

Several unique properties have been found in the iron arsenide superconductors. For example, these materials are semimetals and therefore metallic even without doping, 
in contrast to the cuprates. In BaFe$_2$As$_2$, doping Co into the FeAs-plane also induces superconductivity \cite{Sefat_PRL_78_104505}, which differs from the 
suppression of superconductivity and formation of local moments by any doping into the cuprate CuO-planes.
Superconductivity 
has been reported under hydrostatic pressure in the parent compounds CaFe$_2$As$_2$ \cite{Torikachvili_PRL_101_057006,Park_CM_20_322204,Lee_CM}, SrFe$_2$As$_2$\cite{Igawa_08101377,Kumar,Lonzarich_JPCM}, and 
BaFe$_2$As$_2$ \cite{Lonzarich_JPCM}. In particular, for CaFe$_2$As$_2$, T$_c$ as high as 10K has been found in a moderate 0.4GPa pressure 
\cite{Torikachvili_PRL_101_057006,Park_CM_20_322204,Lee_CM}, while for SrFe$_2$As$_2$ and BaFe$_2$As$_2$, superconductivity is achieved at about 28K at P=3.2~GPa and 4.5~GPa 
respectively \cite{Lonzarich_JPCM}.

In CaFe$_2$As$_2$ in ambient pressure, a structural phase transition (from tetragonal to orthorhombic) is seen at T$_{S1}=170$~K\cite{Ni_prb_78_014523},
accompanied by the appearance of magnetic order\cite{Goldman_PRB_78_100506}; this transition is seen as a sharp upwards anomaly in resistivity. Hydrostatic pressure causes a reduction of T$_{S1}$.
The signature in resistivity becomes a broad upturn, rather than the sharp discontinuous change seen in ambient pressure\cite{Torikachvili_PRL_101_057006,Lee_CM}.
Above 0.5GPa, a collapsed tetragonal structure is 
identified below a separate structural transition temperature (T$_{S2}$) \cite{Torikachvili_PRL_101_057006,Kreyssig_CM}. The collapsed tetragonal phase has the same crystal symmetry 
as the high-temperature one, but with a $\sim10$\% reduction in the c-axis parameter and a 2$\%$ expansion of the in-plane lattice parameters \cite{Kreyssig_CM}.
The transition to the collapsed phase is also seen\cite{Torikachville_CM_08091080} in resistivity measurements, where a broad downward change in slope has been observed. The temperature T$_{S2}$ increases
with further increases in pressure\cite{Torikachvili_PRL_101_057006,Kreyssig_CM}. 
The transition of CaFe$_2$As$_2$ from the high temperature tetragonal 
phase to either of the low temperature phases seems very sensitive to different pressure conditions. Transport studies using a conventional clamp cell see wide
transitions \cite{Torikachvili_PRL_101_057006,Lee_CM}, whereas neutron scattering using helium as the pressure medium suggests that the 
orthorhombic and collapsed tetragonal phases emerge sharply at low temperatures \cite{Kreyssig_CM}.
The maximum superconducting T$_c$ is achieved at about 10K around an inferred phase boundary P$\approx$0.5GPa \cite{Torikachvili_PRL_101_057006,Lee_CM} between the two low temperature
structures. This near vertical boundary was explicitly detected in isothermal pressure sweeps\cite{Kreyssig_CM,Goldman_2008}.
Recently Lee {\em et al.} postulated\cite{Lee_CM} the existence of a third phase in the region of the phase boundary and associated superconductivity with that border phase.

In order to clarify the phase diagram and the nature of the various phase transitions we have studied the high-pressure dc susceptibility and resistivity
of CaFe$_2$As$_2$ using a helium gas pressure system. Compared with clamp 
pressure cells, helium has a low freezing point, which only increases to about 50~K at P$=0.7$~GPa.
CaFe$_2$As$_2$ single crystals were grown by the Sn-flux method \cite{Ni_prb_78_014523}, and afterwards surface Sn was removed by etching with 
HCl. For the transport measurements, samples were loaded in a pressure cell with either a four-probe or a Van der Pauw configuration. The pressure cell was 
cooled in a helium storage dewar and the pressure was applied {\em in situ} by an external helium compressor. For the magnetization measurements using a Quantum Design MPMS, the sample was loaded in a separate 
cell which was connected to the same helium compressor. The maximum 
pressure was about 0.7~GPa for both pressure cells and we employed a cooling rate of about 1~K/min through the structural transitions.

\begin{figure}
\includegraphics[width=9cm, height=9cm]{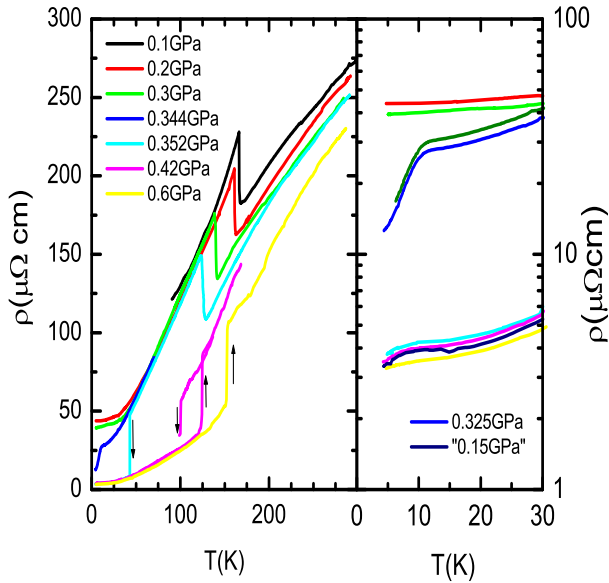}
\caption{\label{Res} (a) The resistivity of a CaFe$_2$As$_2$ single crystal at different pressures. For pressures above 0.35~GPa, the cooling or the warming up direction 
is indicated by an arrow next to the plots. (b) The low-temperature resistivity of the same CaFe$_2$As$_2$ crystal. The 0.15~GPa data is taken after decreasing pressure 
from 0.4~GPa to 0.15~GPa at T=50~K.}
\end{figure}

Fig.~\ref{Res}(a) shows the resistivity of a CaFe$_2$As$_2$ single crystal under different pressures. All measurements were reproducible from sample to sample.
At ambient pressure, the sample shows a sharp increase of resistivity 
at T$_{S1}\approx$170~K, corresponding to a first order structural transition from the tetragonal phase to the orthorohombic/AF phase 
\cite{Ni_prb_78_014523, Goldman_PRB_78_100506}. The residual resistance ratio (RRR) of our sample is about 5-6, which is similar to the other reports of etched 
samples \cite{Ni_prb_78_014523, Torikachvili_PRL_101_057006, Lee_CM}. The resistivity decreases gradually with pressure at room temperature. As pressure increases to 
0.344~GPa, T$_{S1}$ decreases to 120K, and a sharp first order phase transition is still clearly seen. This 
is very different from previous measurements in clamp cells where the resistivity changes gradually with pressure and becomes smooth with temperature at high pressures 
\cite{Torikachvili_PRL_101_057006,Lee_CM}.
Above 0.4~GPa, the transition to the orthorhombic phase, with its sharp
upwards anomaly in resistivity disappears completely.

For P$\geq0.4$~GPa, a second phase transition occurs at about T$_{S2}$=100~K with a sudden drop of resistance as seen in Fig.~\ref{Res}(a), which is known to be from the high-temperature 
tetragonal to the low-temperature collapsed tetragonal structure\cite{Goldman}. T$_{S2}$ increases with increasing pressure, in agreement with
the neutron\cite{Kreyssig_CM} and transport\cite{Torikachvili_PRL_101_057006, Lee_CM} results. Our data show the 
following features of the second phase. i){\it Low resistivity}. From the high-temperature tetragonal phase to the collapsed tetragonal phase, the resistivity drops by a 
factor of two at the transition for all pressures. As seen in Fig.~\ref{Res}(b), the residual resistivity of the collapsed phase is about 3.5~$\mu \Omega$cm, which 
is about about one-tenth that of the low-pressure orthorhombic phase. As originally reported\cite{Torikachvili_PRL_101_057006} the RRR is about 70, which barely changes with pressure
once saturated. ii) {\it 
Sharp transitions}. At all pressures, the resistance drops steeply through the transition with a transition width less than 1~K. This is a striking difference from the liquid
medium measurements\cite{Torikachvili_PRL_101_057006, Lee_CM}. iii) {\it Large thermal hysteresis}.
As shown in Fig.~\ref{Res}(a), the onset temperature of the structural transition changes by 20~K during the cooling and warming up process at P=0.42~GPa, 
which form a hysteresis loop. This is similar to hysteresis reported in ref.\cite{Torikachvili_PRL_101_057006, Lee_CM}. iv) {\it Hysteresis with pressure}. The collapased tetragonal 
phase also shows a strong hysteresis with pressure. As shown in Fig.~\ref{Res}(b), by decreasing the pressure from 0.4~GPa to 0.15~GPa at 50~K, the collapsed tetragonal 
phase is still trapped as indicated by the low resistivity value. The orthorhombic phase is only recovered below 0.1~GPa. This is consistent with scattering measurements taken
in He cells\cite{Kreyssig_CM,Goldman_2008}.

The phase boundary between the orthorhombic phase and the collapsed tetragonal phase is found at P$\approx$0.35~GPa. We did fine tuning of the pressure in steps of 0.01~GPa 
close to 0.35~GPa, and saw a direct transition between the orthorhombic phase and the collapsed tetragonal phase upon cooling at 0.354~GPa. As shown in 
Fig.~\ref{Res}(a), the sample first goes to the orthorhombic phase with a sharp increase of resistivity at T$_{S1}\approx$110~K. On further cooling, there is a direct 
transition from the orthorhombic to the collapsed tetragonal phase with a dramatic decrease of resistivity at T$_{S2}\approx$45~K.

\begin{figure}
\includegraphics[width=9cm, height=9cm]{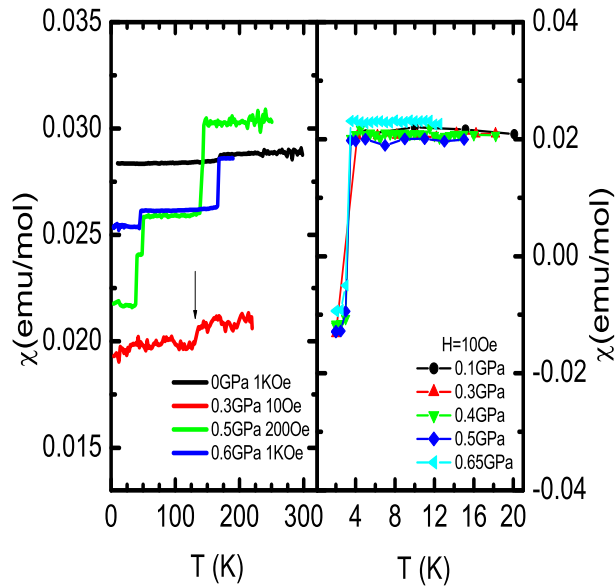}
\caption{\label{Sus}(a) The zero-field-cooled dc susceptibility of CaFe$_2$As$_2$ single crystals under pressure measured in the MPMS. The magnetic field is aligned along 
the crystalline a (or b) axis. (b) The low-temperature dc susceptibility with an applied magnetic field of 10~Oe. }
\end{figure}

Our magnetization data are consistent with the transport results. In Fig.~\ref{Sus} (a), the zero-field cooled (ZFC) dc susceptibility of CaFe$_2$As$_2$ is shown 
for various pressures. At ambient pressure, a drop of susceptibility is clearly seen at the structural transition temperature T$_{S1}\approx $170~K. The transition 
temperature drops to 125~K at P$\approx$0.3~GPa. Further increase of pressure causes another sharp drop of susceptibility at T$\approx$140~K with P$\approx$0.5~GPa, and 
T$\approx$150~K with P$\approx$0.6~GPa, which is consistent with the structural transition T$_{S2}$ seen in resistivity. Below $T_{S2}$, another drop 
of susceptibility is seen at T$\approx$50~K for P=0.5-0.6~GPa. However, this temperature corresponds to the helium solidification temperature at these pressures, 
which suggests that the collapsed tetragonal phase is very sensitive to even the small changes in the pressure environment when the helium freezes and does not represent an additional
phase transition. The susceptibility below 20~K is shown in Fig.~\ref{Sus}(b). 
For all pressures, we did not see any diamagnetism down to 4~K. Below 4~K, there is a diamagnetic signal at all pressures as shown in Fig.~\ref{Sus}(b). We 
ascribe this diamagnetism to superconductivity of very small amounts of unetched tin flux, noting that the superconducting volume corresponds to only about $0.5\%$ of the total sample volume.

\begin{figure}
\includegraphics[width=9cm, height=9cm]{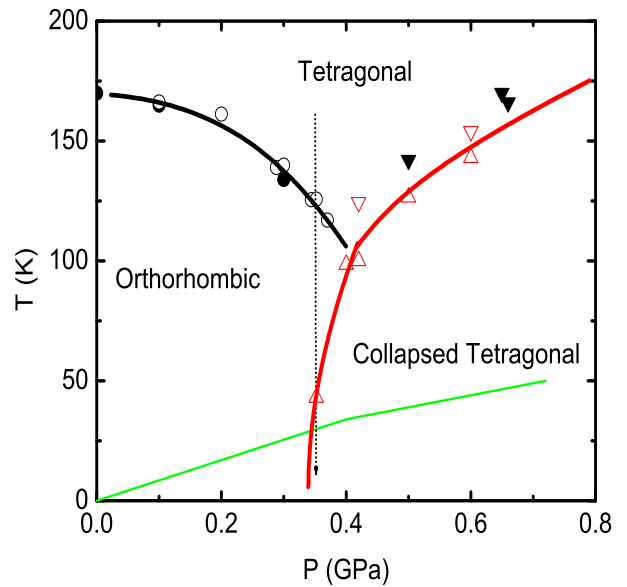}
\caption{\label{PD} The (pressure, temperature) phase diagram of CaFe$_2$As$_2$ constructed from the transport and susceptibility measurements. The solid circles and the 
solid inverted triangles correspond to T$_{S1}$ and T$_{S2}$ respectively measured by susceptibility during warm up after ZFC. The hollow circles and the hollow inverted triangles correspond to
T$_{S1}$ and T$_{S2}$ respectively measured by transport during warm up. The hollow triangles correspond to T$_{S2}$ measured by transport during cooling down. The dashed 
line shows two structural transitions at the pressure P=0.352~GPa. The green line is the helium solidification temperature for reference.}
\end{figure}

In Fig.~\ref{PD}\ we have constructed a (P, T) phase diagram from our transport and susceptibility measurements. The transition temperature measured during 
warm up is consistent for both techniques. For the higher pressure structural transition, the transition temperature first rises quickly with pressure from $P\approx$0.35~GPa to 
0.4~GPa, and then increases by about 20~K/kbar above 0.4~GPa; there is about 20~K of hysteresis in the T$_{S2}$ transition.

Our results are quite different from the transport data using the clamp cells in a few aspects. First, we found a clear phase boundary at P$\approx$0.35~GPa where 
the orthorhombic phase changes to the collapsed tetragonal phase with temperature or pressure abruptly. Second, the higher pressure structural transition is of the first order type, which is 
indicated by the narrow transition width and the hysteresis with pressure and temperature.
Finally and most importantly, we did not see bulk superconductivity at any pressure up to 0.65~GPa.
Careful examination of our transport data at P$\le 0.344$~GPa in Fig.~\ref{Res}(b) shows a drop of resistivity at T$\le$10~K. Since 
superconductivity is reported at T$\approx$10~K by previous transport studies above 0.3~GPa, the drop of resistivity in our case might be caused by some 
superconducting islands. However, any such superconducting regions should be very small since the resistivity is still quite high at 4.2~K and no signature is seen in susceptibility.

A major difference between the the helium pressure cell and the liquid media pressure cells is that helium's solidification temperature is much lower and that even when frozen, helium cannot
support much shear stress. As shown in 
Fig.~\ref{PD}, the helium solidification temperature increases to about 50~K at P$\approx$0.6~GPa, and is much lower than both structural transition temperatures.
At a temperature below the 
helium solidification temperature, as shown in Fig.~\ref{Sus}(a), a sharp drop of susceptibility is clearly seen. In contrast, the solidification temperature of other 
pressure mediums, for example FC-77, is usually much higher than 100~K at P$\approx$0.35~GPa. Therefore it is likely that the pressure is non-hydrostatic through the 
structural transitions around the phase boundary P$\approx$0.35~GPa.

In CaFe$_2$As$_2$ the lattice parameters change dramatically at the transition to the collapsed tetragonal phase\cite{Kreyssig_CM} and so 
pressure homogeneity will be an issue if the sample is embedded in a solid pressure medium at the transition temperature. For example at P$\approx$0.4~GPa, the structural 
transition causes a decrease of the c-axis lattice parameter by $\sim10$\% and an increase in the a-axis parameter by $\sim2$\%. For the worst case, {\it i.e.}, assuming that the frozen pressure medium is unable to make 
plastic adjustment with the volume change of the sample, then the sample clearly must phase separate. 
Such a {\it constant volume} assumption upon cooling, rather than a constant pressure one, seems consistent with 
the $\mu$SR studies using nephane as the pressure medium, where a 50$\%$ volume fraction of the magnetic phase is seen at $P\approx$0.5~GPa \cite{Goko} as well as recent neutron
results\cite{Goldman_2008}.

Non-hydrostatic conditions may cause the formation of domains with different properties (possibly including superconductivity). In a non-hydrostatic condition, large 
domain walls may be generated between highly phase-separated regions. Since the low-temperature orthorhombic and the tetragonal phase have a large lattice mismatch, 
intermediate phases with different lattice parameters could be generated in the domain walls. In particular, if an orthorhombic structure with smaller lattice parameters is 
formed in the domain walls, a virtual high-pressure effect is realized on the orthorhombic phase. This constant volume scenario also suggests that the volume ratio in 
the phase separation region and the pressure range of superconductivity can be different if pressure media with different melting temperatures are used.
It is also possible that superconductivity is caused by a uniaxial component of the pressure.
Uniaxial stress can also cause a constant volume situation, since the uniaxial pressure is not dynamically 
maintained through the transition temperature either. Therefore, it may be hard to distinguish this from the constant volume scenario.

The constant volume scenario does not conflict with the diamagnetic signal from high-pressure ac susceptibility on CaFe$_2$As$_2$ \cite{Lee_CM}. If 
superconductivity is generated in the domain walls to form a thick wall honeycomb-like superconductor, it will be hard to distinguish from 
bulk superconductivity by transport or susceptibility measurements.

In summary, we have studied the high-pressure susceptibility and the transport properties of CaFe$_2$As$_2$, using helium as the pressure medium. Our data have identified 
two first order phase transitions separated at P$\approx$0.35~GPa. In contrast to other high-pressure studies using clamp cells, we did not see any
superconductivity. Therefore, our data indicate that the phase separation and superconductivity in the previous 
studies are most likely caused by a non-hydrostatic component of pressure. Our results invite caution with respect to the nature of
high-pressure superconductivity 
in all three parent compounds, CaFe$_2$As$_2$, SrFe$_2$As$_2$, and BaFe$_2$As$_2$. We note the proposed third phase under pressure \cite{Lee_CM} is not seen in our work. 
Further study is necessary 
to identify the actual phase properties of the superconducting region. 
Local probes, such as NMR or $\mu$SR, should be useful to perform the studies under non-hydrostatic conditions and verify our scenario.

Research at McMaster University is supported by NSERC and CIFAR. Work at Ames Laboratory was supported by the Department of Energy, Basic Energy Sciences under Contract No. DE-AC02-07CH11358.
We appreciate useful discussions with Alan Goldman regarding ref.~\cite{Goldman_2008}.

\end{document}